\documentstyle[12pt]{article}

\catcode`\@=11

\@addtoreset{equation}{section}

\begin{document}

\baselineskip 24pt
\newcommand{\numero}{YCTP-P10-95}
\newcommand{\num}{hep-ph/9509270}
\newcommand{\titre}{AN EXTENDED TECHNICOLOR MODEL WITH}
\newcommand{\titreb}{QCD-LIKE SYMMETRY BREAKING}
\newcommand{\auteura}{Thomas Appelquist  and Nick Evans}
\newcommand{\addressa}{ }
\newcommand{\auteurc}{Tom Appelquist }
\newcommand{\beq}{\begin{equation}}
\newcommand{\eeq}{\end{equation}}
\newcommand{\Fn}{\mbox{$F(p^2,\Sigma)$}}

\input epsf
\newwrite\ffile\global\newcount\figno \global\figno=1
\def\writedefs{\immediate\openout\lfile=labeldefs.tmp \def\writedef##1{%
\immediate\write\lfile{\string\def\string##1\rightbracket}}}
\def\writestoppt{}\def\writedef#1{}

\def\figin{\epsfcheck\figin}\def\figins{\epsfcheck\figins}
\def\epsfcheck{\ifx\epsfbox\UnDeFiNeD
\message{(NO epsf.tex, FIGURES WILL BE IGNORED)}
\gdef\figin##1{\vskip2in}\gdef\figins##1{\hskip.5in}
\else\message{(FIGURES WILL BE INCLUDED)}%
\gdef\figin##1{##1}\gdef\figins##1{##1}\fi}

\def\figinsert{}
\def\ifig#1#2#3{\xdef#1{fig.~\the\figno}
\writedef{#1\leftbracket fig.\noexpand~\the\figno}%
\figinsert\figin{\centerline{#3}}\medskip\centerline{\vbox{\baselineskip12pt
\advance\hsize by -1truein\center\footnotesize{  Fig.~\the\figno.} #2}}
\bigskip\endinsert\global\advance\figno by1}
\def\footnotefont{}\def\endinsert{}

\newcommand{\addressc}{Department Of Physics  \\  Yale  University \\   PO Box
208120\\ New Haven \\CT 06520,USA. }

\newcommand{\abstrait}{We present  a one-doublet extended technicolor model,
with all fermions in fundamental representations.  The bare lagrangian has no
explicit mass terms but generates masses through gauge symmetry breaking by
purely QCD-like dynamics.  The model  generates three families of quarks and
leptons and can accommodate the observed third family mass spectrum (including
a large top mass  and light  neutrinos). In addition, we show how the model may
be extended to incorporate a top color  driven top mass without the need for a
strong U(1) interaction. We discuss the compatiblity of the model with
experimental constraints and its possible predicitive power with respect to
first and second family masses. }

\begin{titlepage}
\hfill \numero \\
$\left. \right.$ \hfill \num
\vspace{.5in}
\begin{center}
{\large{\bf \titre }}
{\large{\bf \titreb}}
\bigskip \\by\bigskip\\ \auteura \bigskip \\ \addressc \\

\renewcommand{\thefootnote}{ }
\vspace{.5 in}
{\bf Abstract}
\end{center}
\abstrait
\end{titlepage}

\def\id{\rlap{1}\hspace{0.15em}1}

\section{Introduction}

The Higgs model of electroweak symmetry (EWS) breaking is less than satisfying
because it offers no understanding of fermion masses and is plagued by a
technical hierarchy problem with respect to the Higgs mass. Technicolor models
\cite{TC} break EWS by the formation of fermion condensates in a strongly
interacting theory patterned after QCD. There are no fundamental scalars and
therefore no Higgs-mass hierarchy problem.  It has been proposed  that the
fermion (quark and lepton) masses could be generated in technicolor models by
extending the gauge sector so that the fermions and the technifermions are
unified above the EWS breaking scale. In such extended technicolor (ETC)
\cite{ETC} models, the hierarchy of fermion masses is generated by a hierarchy
of breaking scales of the unified gauge group. The problem of the origin of the
fermion masses is replaced by the problem of the origin of the ETC symmetry
breaking scales.

 A number of proposals have been made for the origin of the ETC breaking
scales. The ETC symmetries may be broken by including Higgs scalars
\cite{ETChiggs} in appropriate representations of the ETC group. This approach,
however, is usually assumed to be a low energy approximation to an even higher
scale dynamics since it reintroduces the technical hierarchy problem that
technicolor is designed to solve. It  has so far not pointed the way to an
understanding of fermion mass.

A more audacious explanation of the ETC breaking is that the ETC group(s) break
themselves by becoming strong at high scales and forming fermion condensates
which are not singlets under the ETC group. This is the tumbling mechanism
\cite{Tumbling} . It is appealing in its economy,  but  the desired symmetry
breaking patterns require placing the fermions and technifermions in unusual,
non-fundamental representations,  chosen to achieve the desired breaking
pattern.   Furthermore, tumbling models have so far relied on speculative
most-attractive-channel (MAC) analyses to determine the condensates that form
at each scale.

 In this paper, we explore an alternative approach to the ETC symmetry breaking
scales which is purely dynamical (no fundamental scalars and no bare mass terms
in the lagrangian), which puts fermions only in the fundamental representation,
 and which employs only QCD-like dynamics. It thus avoids the use of MAC
analyses as well as  non-fundamental representations. Instead, the breaking
pattern (the pattern of quark and lepton masses) is arranged here by the choice
of groups into which new fermions are placed and the coupling strengths of
these groups. Time will tell whether this holds the key to a deeper
understanding of the quark and lepton masses.

Dynamical models with fermions in only the fundamental representation of the
gauge groups have also been proposed in refs \cite{Moose1, Moose2}. However,
they generate the light fermion masses by means of couplings to new fermions
with mass terms containing the observed mass structure and were intended to
demonstrate that flavor dynamics could be separated from
EW scale physics in ETC models.

The model presented has one doublet of technifermions and  involves Pati-Salam
unification \cite{PS} at high scales. It gives a relatively small contribution
to the electroweak parameter $S$   \cite{PT,Burgess} and gives rise to no
pseudo  Goldstone bosons at the technicolor scale. Within this model we are
able to dynamically generate three family-scales, flavor breaking within each
family, a large top mass, and light neutrinos. The dynamics responsible for
these features do not generate flavour changing neutral currents (FCNC).
FCNCs induced by CKM mixing, the origin of which we do not address here, can be
suppressed by small mixing angles or the familiar  walking \cite{Walk} and
strong ETC \cite{Gap} solutions to the problem.

The model as presented  contains  global symmetries above the highest ETC
breaking scale (typically of order 1000TeV) that, when dynamically broken,
generate exactly massless, physical Goldstone bosons. They couple to ordinary
matter through  ETC interactions or the standard  model  (SM) interactions of
their constituent fermions. These interactions are suppressed by the ETC scale
and are not  visible  in current laboratory experiments. Astrophysical
constraints \cite{redgiant} from stellar lifetimes do, however, rule out light
Goldstones with SM couplings. We anticipate that yet higher scale unifications
than those discussed here may generate masses for these Goldstones which are
above the astophysical constraints.

ETC models that generate the large top mass tend to give rise rise to
contributions to the T \cite{PT,Burgess} parameter that are close to the
experimental bound. The T parameter may be reduced in  top color assisted
technicolor models \cite{Hill} in which the top mass is generated by a close to
critical top self interaction. We show how an alternative model of top color
may be  included simply in our ETC model. Unlike in the orginal top color
model, the isospin breaking that splits the top and bottom masses is the result
of chiral non-abelian color groups rather than a strong $U(1)$ gauge group.

In Section 2, we describe the basic QCD-like mechanism for breaking gauge
symmetries. We apply this dynamics in the case of a  one-doublet model in
Section 3. We discuss both family symmetry breaking leading to different mass
scales for each of the three quark-lepton families, and flavor symmetry
breaking within each family. Phenomenological aspects of the model are also
discussed. In Section 4 we discuss how the model may be extended to include a
variation on  top color assisted technicolor. In Section 5, we summarize the
work and present some conclusions.

\newpage

\section{Gauge Symmetry Breaking With QCD-Like Dynamics}

 In this section, we describe our breaking mechanism using a simple model in
which an $SU(N)$ gauge group is broken to $i$ gauged subgroups and an $SU(j)$
global symmetry group using purely QCD-like dynamics. The driving force is an
additional $SU(M)$ gauge interaction which becomes strongly interacting at a
scale $\Lambda_M$. The model contains the essential dynamics used to break the
ETC symmetries in the following sections. There, the $SU(N)$ group will be the
ETC group, with quarks, leptons, and technifermions in its fundamental
representation. There will also be particles transforming according to the
fundamental representation of both the $SU(N)$ and $SU(M)$ groups, which will
play an active role in the ETC symmetry breaking.  In this section, only the
latter particles will be included
for simplicity.

$\left. \right.$  \hspace{0.3in}\ifig\prtbdiag{}
{\epsfxsize12.8truecm\epsfbox{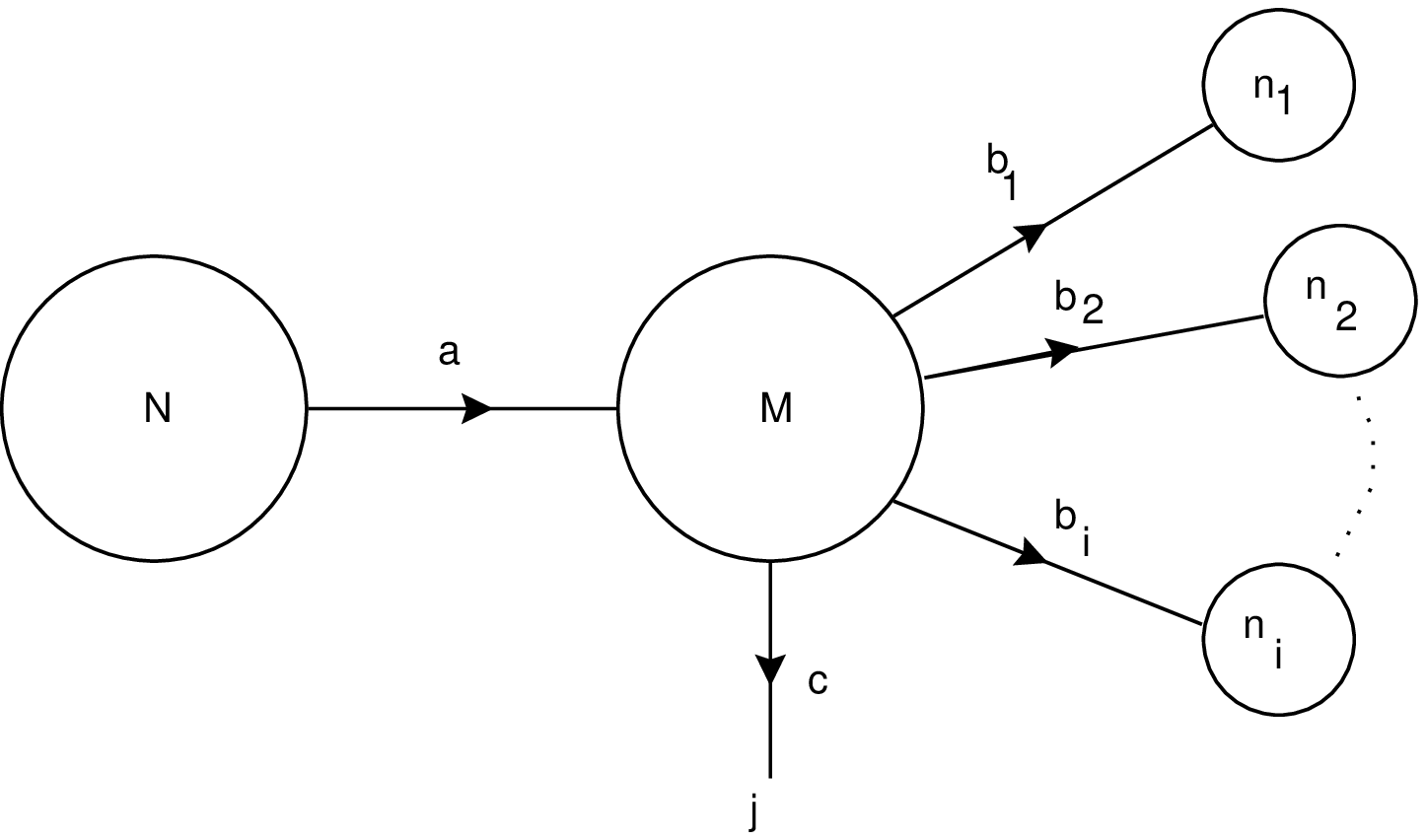}} \vspace{-0.85cm}
\begin{center} Figure 1: A model of gauge symmetry breaking. \end{center}

 In Fig. 1 we show the model in moose notation \cite{Moose1} with

\beq n_1 + n_2 +... +n_i + j = N. \eeq

A circled number $N$ corresponds to an $SU(N)$ gauge symmetry and directional
lines represent left-handed Weyl fermions that transform according to the
fundamental representation of the gauge groups they connect. A line leaving
(entering) a circle with a number $N$ inside represents a fermion transforming
under the $N$ ($\bar{N}$) representation of that group.  Lines labelled by a
number $j$ that is not circled correspond to $j$ copies of the representation
of the gauge group and hence have a global symmetry $SU(j)\otimes U(1)$.

The fermion content of the model pictured in Fig 1 is therefore

\beq \begin{array}{ccccccc}
 & SU(N) & SU(M) & SU(n_1) & SU(n_2) & .... & SU(n_i)\\
&&&&&&\\
a &  N & \bar{M} & 0 & 0 & .... & 0\\
&&&&&&\\
b_1&  0 &  M  & n_1 & 0 & .... & 0\\
&&&&&&\\
b_2 &  0 &  M  & 0 & n_2 & .... & 0\\
&&&&&&\\
:  &  : & : & : & : & : & :\\
&&&&&&\\
 b_i&  0 &  M  & 0 & 0 & .... & n_i\\
&&&&&&\\
 c &  0 &  M  & 0 & 0 & .... & 0
\end{array} \eeq

\noindent where the index a runs over the $j$ flavours  of the c-fermions. This
model is not anomaly free as shown but we shall assume that the additional
degrees of freedom required to make $SU(N)$ and the $SU(n_i)$ gauge groups
anomaly free do not transform under the $SU(M)$, which is anomaly free with the
constraint of Eq. (2.1). The $SU(M)$ will be the only strongly interacting
gauge group at its confinement scale $\Lambda_M$.

  At this scale, the confining $SU(M)$ interaction leads to the formation of
the condensates

\beq <\bar{a}^{1..n_1} b_1> \neq 0, \hspace{.5cm} <\bar{a}^{n_1+1...n_1+n_2}
b_2> \neq 0, \hspace{.5cm} .... ,\hspace{.5cm} <\bar{a}^{n_1+n_2+...+n_i+1
...N} c> \neq 0. \eeq

\noindent  With the other gauge interactions neglected, the global symmetry on
the fermions $a$, $b$ and $c$, would be
$SU(N)_L \otimes SU(N)_R$. The condensates break this symmetry in the usual
pattern

\beq SU(N)_L \otimes SU(N)_R \rightarrow SU(N)_V. \eeq

\noindent  In the presence of the other gauge interactions, the gauged $SU(N)$
group is therefore broken to

\beq  SU(n_1) \otimes SU(n_2) \otimes.... \otimes SU(n_i),  \eeq

\noindent where the gauge field and gauge coupling for each group is a linear
combination of the fields and couplings of Fig. 1.  We note that all $N^2-1$
Goldstone bosons associated with the broken symmetry are eaten by the $N^2-1$
gauge bosons that acquire a mass (of order $\Lambda_M$).

  This symmetry breaking mechanism is of course reminiscent of technicolor
itself. Here, as there, the symmetry breaking is driven by an additional,
strongly coupled gauge interaction, and the breaking  pattern is being imposed
by the choice of the $SU(n_i)$ gauge groups.  In each case, this is to be
compared with the choice of scalar representation in the Higgs mechanism. For
ETC breaking, it can also be compared with the choice of fermion
representations in tumbling models.

\newpage

 \section{One Doublet Technicolor}

   As an example of ETC breaking using the above mechanism, we construct an ETC
model with a single doublet of technifermions, $U$ and $D$ \cite{TC,Moose2}

\beq Q_L = \left(\begin{array}{c} U \\ D \end{array} \right)_L, \hspace{0.2in}
Q_R = \left(\begin{array}{c} U \\ D \end{array} \right)_R .\eeq
The quarks and leptons  must be unified in a single ETC multiplet with the
technifermion doublet. The simplest realization of this unification is a Pati
Salam \cite{PS} $SU(N+12)$ symmetry where the technicolor group is $SU(N)_{TC}$
and where the SM fermions and technidoublet form the multiplets

\beq \begin{array}{l}
 {\cal U}_R = ( U, t,\nu_{\tau}, c, \nu_{\mu}, u, \nu_e )_R,\\
\\
\Psi_L = \left( \left( \begin{array}{c}U\\D\end{array} \right),
\left( \begin{array}{c}t \\ b\end{array} \right),
\left( \begin{array}{c} \nu_{\tau}\\ \tau\end{array} \right),
\left( \begin{array}{c} c \\ s \end{array} \right),
\left( \begin{array}{c} \nu_{\mu} \\ \mu \end{array} \right),
\left( \begin{array}{c} u \\ d \end{array} \right),
\left( \begin{array}{c} \nu_e \\ e \end{array} \right) \right)_L,\\
\\
{\cal D}_R = ( D, b,\tau, s, \mu, d, e)_R. \end{array} \eeq

\subsection{Family Structure}

\subsubsection{A Single Family Model}

To introduce the model we restrict attention to the technidoublet and the third
family quark and leptons only. The ETC group is then $SU(N+4)$. The model is
shown in moose notation in Fig 2.

$\left. \right.$ \hspace{1cm}
\ifig\prtbdiag{}
{\epsfxsize16.8truecm\epsfbox{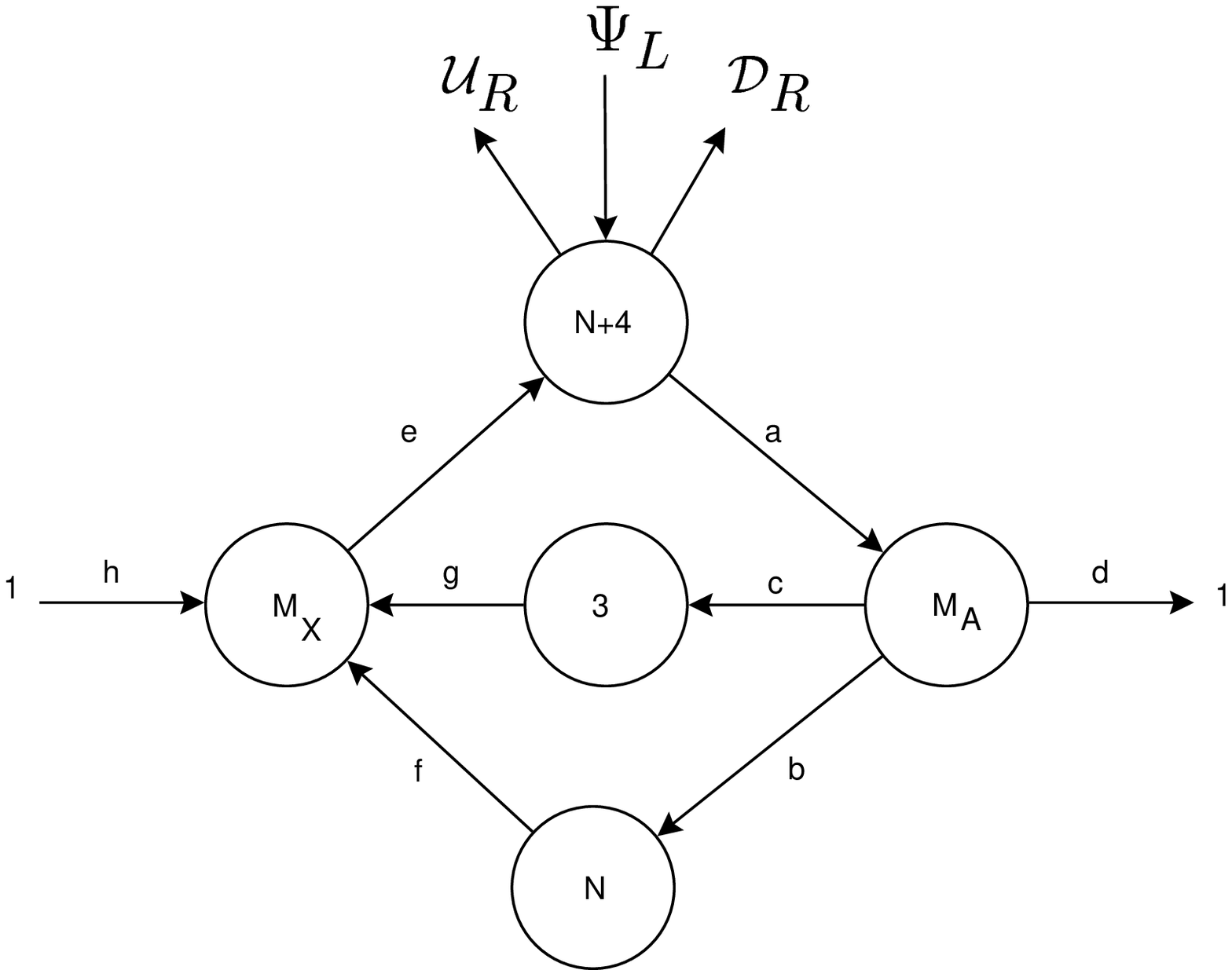}} \vspace{-1.85cm}
\begin{center} Figure 2: A one family, one technidoublet ETC model .
\end{center}

The $SU(M)$ gauge groups become strong in the order A and then X (at  scales
$\Lambda_A$ and $\Lambda_X$ both of order a few TeV), triggering the breaking
of the ETC group to $SU(N)_{TC}$. Consider the highest of these two scales,
$\Lambda_A$. The fermions transforming under $SU(M)_A$ also transform according
to the fundamental representations of  the gauged $SU(N+4)\otimes SU(N)\otimes
SU(3)$. The $SU(3)$ gauge group is present in order to leave an unbroken
$SU(3)$ subgroup of $SU(N+4)$, which will become QCD, acting on the third
family of quarks. The strong $SU(M)_A$ interactions form condensates
\vspace{0.3in}

\beq <\bar{a}^{1..N} b> \neq 0, \hspace{1cm}  <\bar{a}^{ N+1..N+3} c> \neq 0,
\hspace{1cm}  <\bar{a}^{N+4} d> \neq 0, \eeq

\noindent breaking the gauged  $SU(N+4)\otimes SU(N)\otimes SU(3)$ symmetry to
$SU(N) \otimes SU(3)_{QCD}$. The multiplets in Eq. (3.2) are broken, with the
SU(3) subgroup corresponding to the $t$ and $b$ quarks with QCD interactions,
the singlet to the first family lepton doublet and the $SU(N)$ subgroup  to the
unbroken technicolor gauge group.  All $(N+4)^2-1$ Goldstone bosons generated
at this first stage of breaking are eaten by gauge bosons which acquire masses
of order the confinement scale.

The ETC gauge bosons corresponding to  generators broken at $\Lambda_A$ acquire
masses of order $F_A$ , the decay constant of the Goldstone bosons formed at
$\Lambda_A$ that are eaten by the gauge bosons ($F_A^2\simeq M
\Lambda_A^2/4\pi^2$). Below the technicolor scale, where the technifermions
condense, these gauge bosons will generate masses for the third family quarks
and leptons given by

\beq m_f \simeq {\langle \bar{Q}Q \rangle \over F_A^2}, \eeq
where we have assumed that the ETC coupling is perturbative and have used  the
four fermion approximation for the ETC gauge boson.The ETC gauge boson's mass
is proportional to its coupling ($M_{ETC}^2 \simeq g^2F_A^2$) and hence the ETC
coupling cancels in the quark and lepton masses.
In this simple model the quarks and leptons are degenerate. We shall address
generating flavor breaking within each family in section 3.2.

To cancel anomalies in the model, the additional fermions, $e$, $f$, $g$ and
$h$, transforming under the $SU(M)_X$ gauge group have been introduced. The
$SU(M)_X$ group confines these new fermions to remove them from the physical
spectrum at low energies. We assume that this confinement scale, $\Lambda_X$,
lies between the technicolor scale and the  $SU(M)_A$ confinement scale.  At
the scale $\Lambda_X$ there is a global $SU(N+4)_L \otimes SU(N+4)_R$ symmetry
acting on the fermions transforming under $SU(M)_X$.The prefered vacuum
alignment is that no gauge interactions are broken at this extra breaking scale
so there are $(N+4)^2-1$  Goldstone bosons which are not eaten.  The
Goldstone's that transform under the adjoint or fundamental representations of
technicolor or QCD  acquire masses governed by the scale $\Lambda_X$. The
remaining two  Goldstones are massless and we leave discussion of them to
section 3.7.

$\left. \right.$ \hspace{0.15in}
\ifig\prtbdiag{}
{\epsfxsize12.8truecm\epsfbox{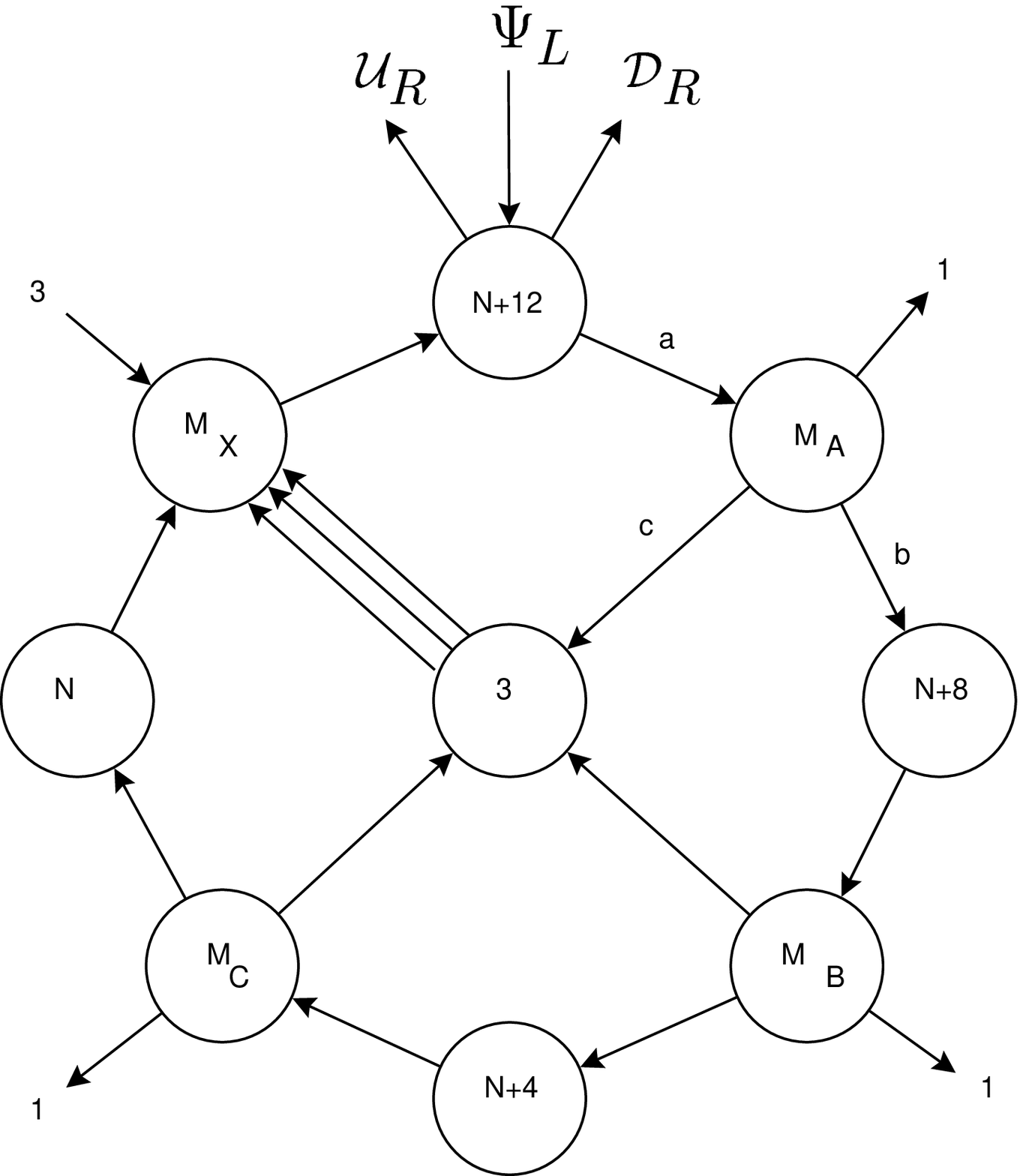}} \vspace{-0.65cm}
\begin{center} Figure 3: A one doublet ETC model with  three family scales.
\end{center}

\subsubsection{Three Families}

The model can be generalized to include three families of quarks and leptons as
shown in Fig 3. The  ETC symmetry $SU(N+12)$ is broken  to $SU(N)_{TC} \otimes
SU(3)_{QCD}$ at three separate scales. There is a separate SU(M) group to
trigger the breaking at each scale. Each is assumed to become strongly
interacting in the order A (at a scale of order a few 100's of TeV), B (at a
scale of order a few 10's of TeV), and finally C (at a scale of order a few
TeV). At each scale the breaking pattern is the same as that discussed in the
one family model; at $\Lambda_A$ the ETC symmetry $SU(N+12)$ is broken to
$SU(N+8) \otimes SU(3)$.
This breaking pattern is then repeated by the groups B and C. At the scale
$\Lambda_B$  it is the SU(3) containing the SU(3) subgroup of SU(N+12)  broken
at the scale $\Lambda_A$,  and an SU(3) subgroup of the SU(N+8) group that
break together to an SU(3) group that finally at the lowest breaking scale
becomes QCD. The QCD interactions are finally shared by all quarks in the
model.  The broken gauge bosons of the ETC group now divide into three sets:
those with masses of order $F_A$ connecting the first family of SM fermions to
more massive generations; those with masses of order $F_B$ connecting the
second family  to more massive generations; and those with masses of order
$F_C$ connecting the third family   to technifermions. This hierarchy of ETC
gauge bosons masses will generate the hierarchy of
quark and lepton family masses below the technicolor scale.

Anomalies are again cancelled in the model by the fermions transforming under
the extra $SU(M)_X$ gauge group that confines these fermions between the
technicolor and lowest ETC scale. In the enlarged model there are 6 Goldstone
bosons that have no gauge interactions and are hence massless.

\subsection{Flavor Symmetry Breaking}

The model in Fig 3  has an $SU(8)$ flavor symmetry within each family, broken
only by the weak SM interactions. To generate the observed quark and lepton
masses we must introduce quark-lepton symmetry breaking interactions and
isospin symmetry breaking interactions for both the quarks and leptons. For
ease of understanding let us discuss a model of just the third family and the
technidoublet.
\vspace{-0.5cm}

\subsubsection{Isospin Breaking}

We shall break isospin degeneracy by  making the ETC gauge group chiral
\cite{Chiral}.
We take it to be  $SU(N+4)_L \otimes SU(N+4)_{{\cal U}_R}\otimes SU(N+4)_{{\cal
D}_R}$, as shown in the model in Fig 4. The one family model in Fig 2 is shown
by the full lines in Fig 4, with the additional sectors discussed in this
section shown as dashed lines.

$\left. \right.$ \hspace{-0.35in}
\ifig\prtbdiag{}
{\epsfxsize15truecm\epsfbox{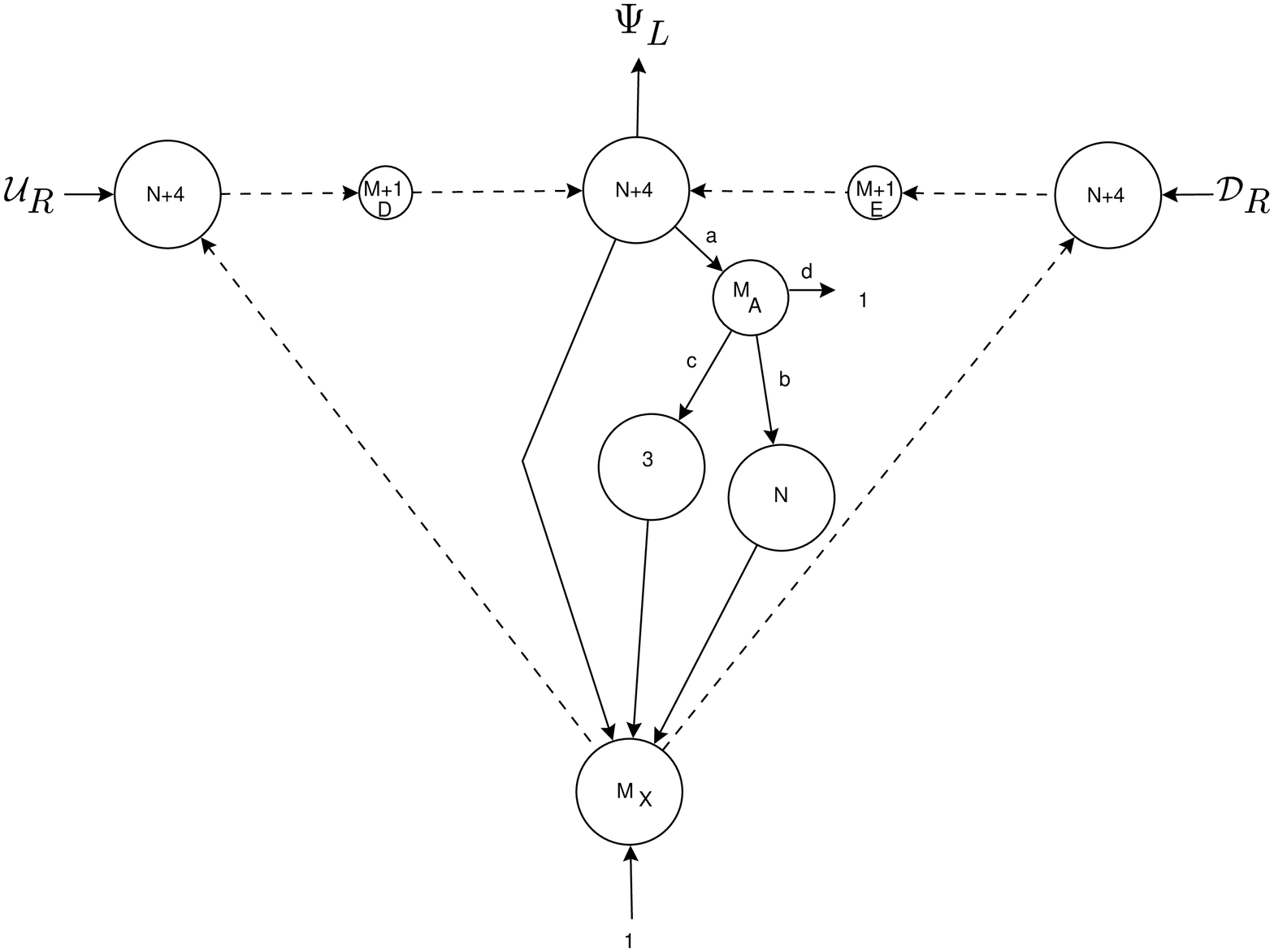}} \vspace{-1.55cm}
\begin{center} Figure 4: Isospin breaking in the model of the third family.
\end{center}

The $SU(M)_A$ gauge group forms condensates at  $\Lambda_A$  and breaks the
$SU(N+4)_L$ ETC group to $SU(N)\otimes SU(3)$ as in the simplest model. The two
gauge groups $SU(M+1)_{D}$ and $SU(M+1)_E$ then become strong between the scale
$\Lambda_A$ and the technicolor scale (for the purposes of making estimates we
shall take $\Lambda_A \simeq \Lambda_D \simeq \Lambda_E $), breaking the chiral
ETC groups to the vector $SU(N)_{TC} \otimes SU(3)_{QCD}$. At each of these
breakings, all Goldstone modes are eaten by gauge bosons associated with broken
generators.

There are now three degrees of coupling freedom associated with the
interactions of the quarks and leptons: the  $SU(N+4)_L$ coupling  $g_L$; the
$SU(N+4)_{ {\cal U}_R}$ coupling $g_{{\cal U}_R}$; and the $SU(N+4)_{{\cal
D}_R}$ coupling $g_{{\cal D}_R}$. The couplings that enter into the quark and
lepton masses are these running couplings evaluated at the breaking scale of
the ETC interactions and they  will in general  break the isospin symmetry of
the model. The left and right handed ETC gauge bosons mix through loops of the
fermions transforming under $SU(M+1)_D$ and $SU(M+1)_E$ that have condensed at
$\Lambda_{D,E}$ as shown in Fig 5.
We shall use these extra degrees of freedom to generate the top-bottom mass
splitting. The two extra parameters will not be sufficient to explain
quark-lepton mass differences which we leave to the next section.

$\left. \right.$ \hspace{-0.35in}
\ifig\prtbdiag{}
{\epsfxsize7.8truecm\epsfbox{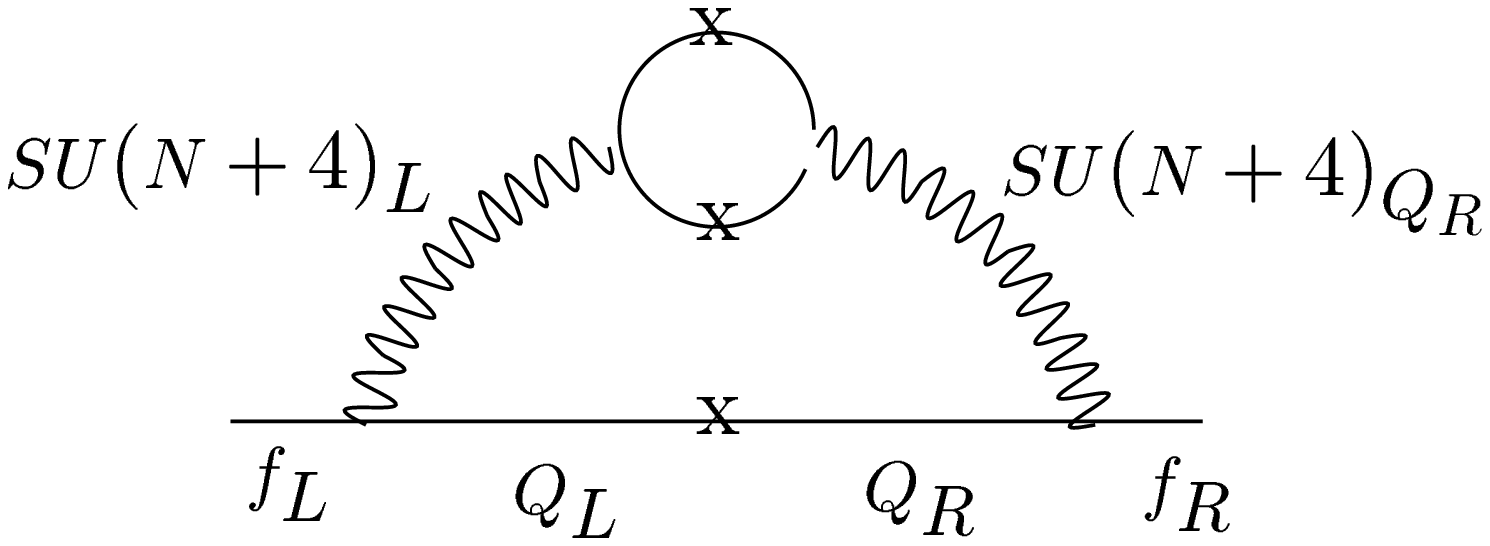}} \vspace{-1.55cm}
\begin{center} Figure 5: Generation of third family fermion, $f$, mass from the
technifermion, $Q$, condensate. \end{center}

If we assume that the ETC gauge bosons coupling to the top have couplings,
$g_L$ and $g_{{\cal U}_R}$, of order one or greater (but less than $4\pi$ at
which the ETC gauge bosons become strongly coupled) these gauge bosons will
have masses of order $F_{A} \simeq F_D$ or larger.We may  approximate them at
the technicolor scale by four fermion interactions.
The ETC couplings cancel as in Eq(3.4) and the top mass can be estimated to be
roughly
\beq m_t \simeq {N \over 4\pi^2} {\Sigma(0)^2 \over F_A^2} \Sigma(0) \eeq
where $\Sigma(p)$ is the dynamical technifermion mass. A simple Pagels-Stokar
\cite{Pagels} estimate, compatible with QCD, gives $v^2 \equiv (250GeV)^2
\simeq \frac{N}{4 \pi^2} \Sigma(0)^2$ and hence $\Sigma(0) \simeq 1TeV$ for $N
\simeq 2$.To generate  $m_t$ in the $100+GeV$ range  therefore requires $F_A
\simeq 800GeV$. Although $F_A$ must be close to the technicolor scale,  the
scale $\Lambda_A \simeq 2\pi F_A/\sqrt{M}$ will be larger, as in QCD, and hence
there is some running space for the technicolor coupling to evolve from its
value at the breaking scale to its critical value at the technicolor scale. The
estimates above are clearly naive approximations to the full non-perturbative
technicolor dynamics   and are not to be trusted to more than  factors of two.
It is therefore not completely clear whether a $175GeV$ top mass may be
generated by  perturbative ETC gauge bosons.

If the ETC coupling is raised close to its critical value (the value of the ETC
coupling at which the ETC interactions alone would break the chiral symmetry of
the quark and leptons) at the ETC breaking scale then the approximations above
are not valid and the ETC coupling will not cancel from the top mass.  A
$175GeV$ top mass may  be generated,  though how close the ETC coupling must be
to its critical coupling is unclear. We shall assume that the ETC interactions
are perturbative in the discussions below leaving the possibility that they
might be strong and near-critical to section 4.

To generate a smaller bottom mass we take the coupling $g_{{\cal D}_R}$ to be
less than one. The ETC gauge bosons associated with $SU(N+4)_{{\cal D}_R}$
therefore acquire a mass $g_{{\cal D}_R} F_E$, which are light relative to
$F_{D,E} \simeq F_A \simeq \Sigma(0)$. Referring again to Fig 5, the bottom
mass is given approximately by
\beq m_b \simeq {N \over 4 \pi^2} \int dk^2 k^2 {\Sigma(k) \over k^2 +
\Sigma(k)^2} {g_{{\cal D}_R}^2 \over k^2 + g_{{\cal D}_R}^2 F_A^2} \eeq
where we have taken $F_E = F_A$ and set the external momentum to zero. With
$g_{{\cal D}_R}^2 F_A^2 < \Sigma(0)^2$, the integral can be estimated to give
roughly.
\beq m_b \simeq {N \over 4 \pi^2} g_{{\cal D}_R}^2 \Sigma(0)  \eeq
where we have again  neglected interactions between the ETC gauge boson and the
technicolor gauge bosons. The bottom mass  is thus  suppressed relative to the
top mass by $g_{{\cal D}_R}^2$. The choice $g_{{\cal D}_R} \simeq 1/6$ gives a
realistic value for $m_b$ and leads to a mass of order 200-300GeV for the
$SU(N+4)_{{\cal D}_R}$ ETC gauge boson.

The technifermion mass splitting $\Delta\Sigma(p) \equiv \Sigma_U(p) -
\Sigma_D(p)$ can also be estimated perturbatively in the ETC interactions. The
main contribution in the model is from the isospin violating, massive gauge
bosons that transform under the adjoint representation of SU(N). The splitting
can be estimated to be roughly
\beq \Delta \Sigma \simeq  {N \over 4\pi^2} {\Sigma(0)^3 \over F_A^2} \simeq
m_t . \eeq
\noindent We discuss the implications of this mass splitting for the $\Delta
\rho \equiv \alpha T$ parameter  in section 3.5 below.
\vspace{4cm}

$\left. \right.$ \hspace{-0.35in}
\ifig\prtbdiag{}
{\epsfxsize15 truecm\epsfbox{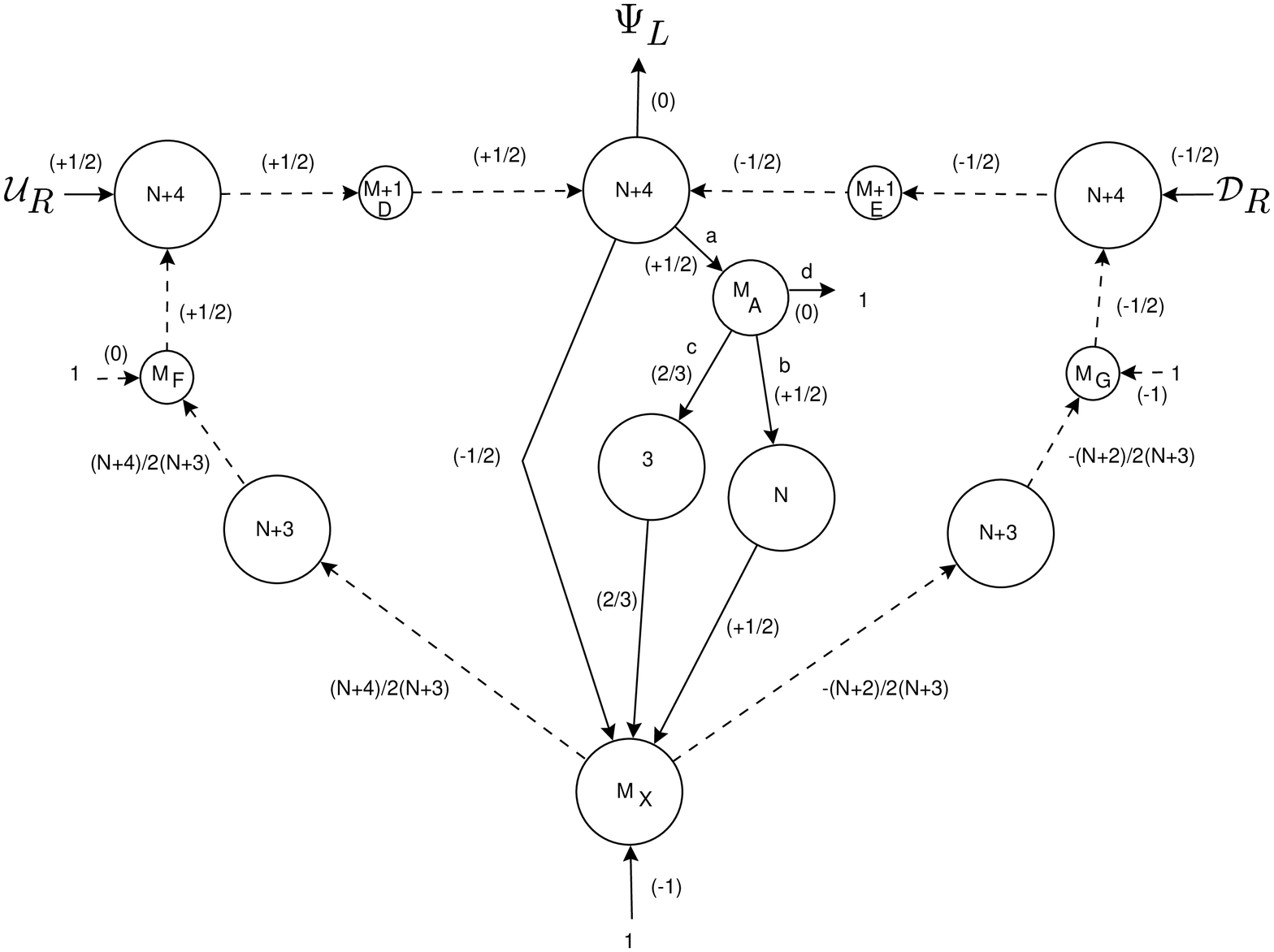}} \vspace{-1.55cm}
\begin{center} Figure 6: Quark lepton mass splitting in the model of the third
family. The fermion lines are labelled by their $U(1)$ hypercharges.
\end{center}

\subsubsection{Lepton Masses}

In the model in Fig 4, the  lepton's interactions are only split from their
quark isospin partners by SM interactions. Although QCD interactions may be
enhanced if the ETC interactions are close to critical (a possibility we
discuss below) and hence could possibly explain the tau-bottom mass splitting
they can not explain why the tau neutrino is so light or massless. In order to
give a fully perturbative ETC model we shall generate the tau-bottom and tau
neutrino-top mass splittings by further ETC breaking dynamics at new scales.

The extra sectors are shown in Fig 6.  The $SU(M)_F$ gauge group becomes
strongly interacting at the scale $\Lambda_F$ and breaks a single gauge color
from the $SU(N+4)_{{\cal U}_R}$ gauge group. The corresponding broken
eigenstate of the multiplet in (3.2) will become the neutrino with mass
\beq m_{\nu_\tau} \simeq {N \over 4\pi} {F_D^2 \over F_A^2}
{\Sigma(0)^3 \over F_F^2} \eeq
with $F_A \sim F_D$ and with a suitably high choice of $F_F$ ($\geq 100TeV$)
the tau neutrino mass may be suppressed below the experimental bound of roughly
$30MeV$.

The gauge group $SU(M)_G$ plays the same role for the tau lepton suppressing
it's mass relative to the bottom quark's by $F_E^2/ F_G^2$ from which we learn
that  $F_G $ must  be of order a few TeV to reproduce the observed tau-bottom
mass splitting.

\subsection{The First And Second Families}

The lightest two families of quarks and leptons may be incorporated in the
model following the discussion in section 3.1.2 and will have mass scales set
by the higher two ETC breaking scales. The top-bottom mass splitting will feed
down to the lightest two family quarks, generating isospin breaking that could
explain the charm-strange mass splitting. The three right handed neutrinos
could all be broken from their ETC multiplet at the scale $\Lambda_F$. The
neutrino masses would then be suppressed relative to the charged lepton masses
by $(F_D/F_F)^2$. The single scale $\Lambda_F$ could thus serve to explain the
lightness of all three neutrinos.The quark-lepton mass splittings  however can
probably  not be generated from the third family in perturbative ETC models,
since the bottom and tau contributions to, for example, the strange and muon
masses are small in comparison with the feeddown from the technifermions' self
energies. If neccessary extra breaking scales may  be introduced to explain the
splittings  using the dynamics discussed above. Similarly the  ETC gauge groups
acting on the right handed up and down quarks may be broken at additional
scales  providing the freedom to accommodate the up-down mass inversion.The
symmetry breaking patterns presented here are not capable of producing the CKM
mixing angles in the quark sector since the families correspond to distinct ETC
gauge eigenstates broken at different scales. We leave the generation of the
mixing angles for future work.

\subsection{U(1) Embedding}

Hypercharge may be embedded in the moose model of Fig 6 by assigning each
particle the  $U(1)$ charge indicated on the fermionic lines. The final
hypercharge group is a subgroup of  the $U(1)_R$ group of the quarks, leptons
and technifermions and the broken diagonal generators of  the $SU(N+4)$ ETC
group. To achieve the correct  breaking pattern the condensates formed by
$SU(M)_A$ must be invariant to $U(1)_Y$. Since the $SU(N+4)$ symmetry of the
fermions transforming as an $\bar{M_A}$ is explicitly broken their $U(1)$
charges must correspond to the relevant subgroup of their $SU(N+4)\otimes U(1)$
symmetry.

\subsection{Phenomenology}

Since there is only one technidoublet in the model  there are  no pseudo
Goldstone bosons generated at the technicolor scale. The single doublet will
also generate only  a small
contribution to the $S$ parameter
 \cite{PT,Burgess}, $S \sim  0.1N$, which we expect to be compatible with the
current experimental two standard deviation upper limit $S < 0.4$.

The isospin violating ETC interactions
will, of course,  give rise to a contribution to the $\Delta \rho (= \alpha T)$
parameter.
The W and Z masses are generated by  techifermion condensation and deviations
from the
$\Delta \rho$ parameter from corrections to the relevant diagrams due to
exchange of isospin violating ETC gauge bosons. At first order in the ETC
interactions the largest contribution will be generated by the exchange of the
massive  gauge bosons transforming under the adjoint of $SU(N)$ across the
techifermion loop. We estimate this  ``direct" contribution \cite{Tom} to be
\beq \Delta \rho \simeq {v^2 \over  8 F_A^2}\eeq
which is of order a  percent.

The isospin violation of the ETC interactions will also feed into the
technidoublet giving rise to mass splitting between the techniup and technidown
(estimated in Eq(3.8)). There is thus an ``indirect" contribution to the
$\Delta \rho$ parameter from loops of non-degenerate technifermions which is
second order in ETC gauge boson exchange. Roughly estimating the contribution
using the peturbative result for $\Delta \rho$ \cite{PT} and the estimate of
$\Delta \Sigma$ in Eq(3.8) we find

\beq  \Delta \rho \simeq {N \Delta \Sigma^2 \over 12 \pi^2 v^2} \simeq { v^4
\over 3 F_A^4 }.\eeq
\noindent These estimates of $\Delta \rho$ are of course naive, ignoring the
effects of the strong technicolor dynamics between the technifermion loops and
neglecting a complete analysis of the many massive ETC gauge bosons. If they
are accurate they could be difficult to  reconcile with the experimental
constraint $\Delta \rho {\
\lower-1.2pt\vbox{\hbox{\rlap{$<$}\lower5pt\vbox{\hbox{$\sim$}}}}\ } 0.3\%$. We
leave a more detailed computation of  $\Delta \rho$ to a subsequent paper.
In any case, in section 4 below we present a variation of the model that will
not overly infect $\Delta \rho$.

The model will also give rise to corrections to the $Zb\bar{b}$ vertex . These
arise from both the exchange of the {\it sideways} gauge boson \cite{Zbb},
coupling technifermions to the bottom, across the $Zb{\bar b}$ vertex and from
mixing of the Z with the diagonal broken ETC generator \cite{Wu}. Each of these
contributions can be as large as a few percent for an ETC scale of order 1TeV
but have opposite signs. The magnitude and sign of the combined correction has
been shown to be compatible with the experimental measurement for some models
(the exact correction is dependent on $N$ and the relative sizes of $g_L$ and
$g_{{\cal U}_R}$).

As presented the model does not give rise to quark or lepton number changing
FCNCs since each family's quark and lepton number are conserved ETC charges in
the model. Of course the most stringent FCNC constraints on ETC models come
from $K^0  {\bar K}^0$ mixing through the CKM mixing angles which break quark
number within each family to a single subgroup. Since we have not addressed the
generation of these mixing angles in this paper we can not address this
constraint. We note though that these FCNCs may be suppressed in several ways;
by small mixing angles in the up-type quark sector, or by a walking technicolor
theory or strong ETC interactions that enhance the ETC scales.

\subsection{Massless Goldstone Bosons}

Massless Goldstone bosons are generated in the model at the scale $\Lambda_X$
as discussed above. These Goldstones carry no charge under any of the gauge
groups in the model. However, their constituents are charged, so they can be
produced by gluon or photon fusion or in the decay of the $Z$ \cite{PGB}. They
can also be produced through the exchange of the heavy ETC gauge bosons. The
amplitude in each case is proportional to $1/F_X$ where $F_X \simeq 1TeV$, so
that the production rate is down by at least an order of magnitude relative to
the production of the Goldstones composed of technifermions that arise in a one
family technicolor model. The rate is below current laboratory limits. With the
Goldstones massless or very light, however, their production by the above
mechanisms is a  major energy loss mechanism for stars \cite{redgiant}, and is
ruled out by stellar abundances.

The Goldstones are thus  troublesome but may acquire masses from further
unifications above the scales discussed in the model so far. In the spontaneous
breaking at $\Lambda_X$, the Goldstone bosons complete an adjoint
representation of the unbroken  $SU(N+12)$ vector global symmetry group  (in
the three family model). If at some higher scale this group is gauged
(corresponding for example to gauging  the full chiral symmetry group in Fig 3)
 then all  the Goldstones will acquire masses given by
\beq M_A^2 \simeq 4\pi F_{X}^4/\Lambda_{new}^2\eeq
 which is potentially sufficient to ensure that the Goldstones will not be a
source of energy loss in stellar interiors.

\section{Strong ETC and Chiral Top Color}

The  model presented so far appears capable of producing a 175GeV top mass
treating the ETC interactions perturbatively without contradicting other
experimental bounds. However, the contributions of the isospin violating ETC
gauge bosons to the  $\Delta \rho$ parameter and  to the $Zb{\bar b}$ vertex
are close to experimental limits. These contributions, which scale as $1/
M_{ETC}^2$, can be reduced by increasing the lowest ETC scale,  but at the
expense of tuning the ETC coupling close to it's critical value from below to
generate the large top mass.   A near critical ETC interaction for the third
family would also enhance  the QCD corrections to the third family quark masses
and potentially explain the bottom tau mass splitting without the need for the
extra ETC symmetry breaking scale $\Lambda_G$ discussed in section 3.2.2.
Finally increasing the lowest ETC scale would allow us to increase the scale
$\Lambda_X$ and hence generate larger masses for the Goldstones formed at that
scale.

Although near critical ETC interactions at a higher ETC scale may suppress the
direct contribution to  $\Delta \rho$ the indirect contribution will remain
roughly the same size but may no longer be considered second order. This
follows from a gap equation analysis which suggests that the technifermion mass
splitting will remain of order $m_t$. Therefore if the large top mass is the
result of either  perturbative or strongly interacting sideways ETC
interactions the contribution to the  $\Delta \rho$ parameter may conflict with
the experimental limit.

Recently Hill \cite{Hill} has proposed that the large top mass may be generated
by a near critical top self interaction \cite{Topcond}. If the ETC gauge boson
with the large isospin violating coupling responsible for the top mass  does
not couple to the technifermions then the isospin splitting will not feed back
into the technisector and hence the $\Delta \rho$ parameter as described above.
Hill generates the top self interaction by assuming that at ETC scales there is
a separate $SU(3)_C \otimes U(1)_Y$
gauge group acting on the third family that is near critical when broken to the
SM gauge groups.

$\left. \right.$ \hspace{-0.35in}
\ifig\prtbdiag{}
{\epsfxsize15 truecm\epsfbox{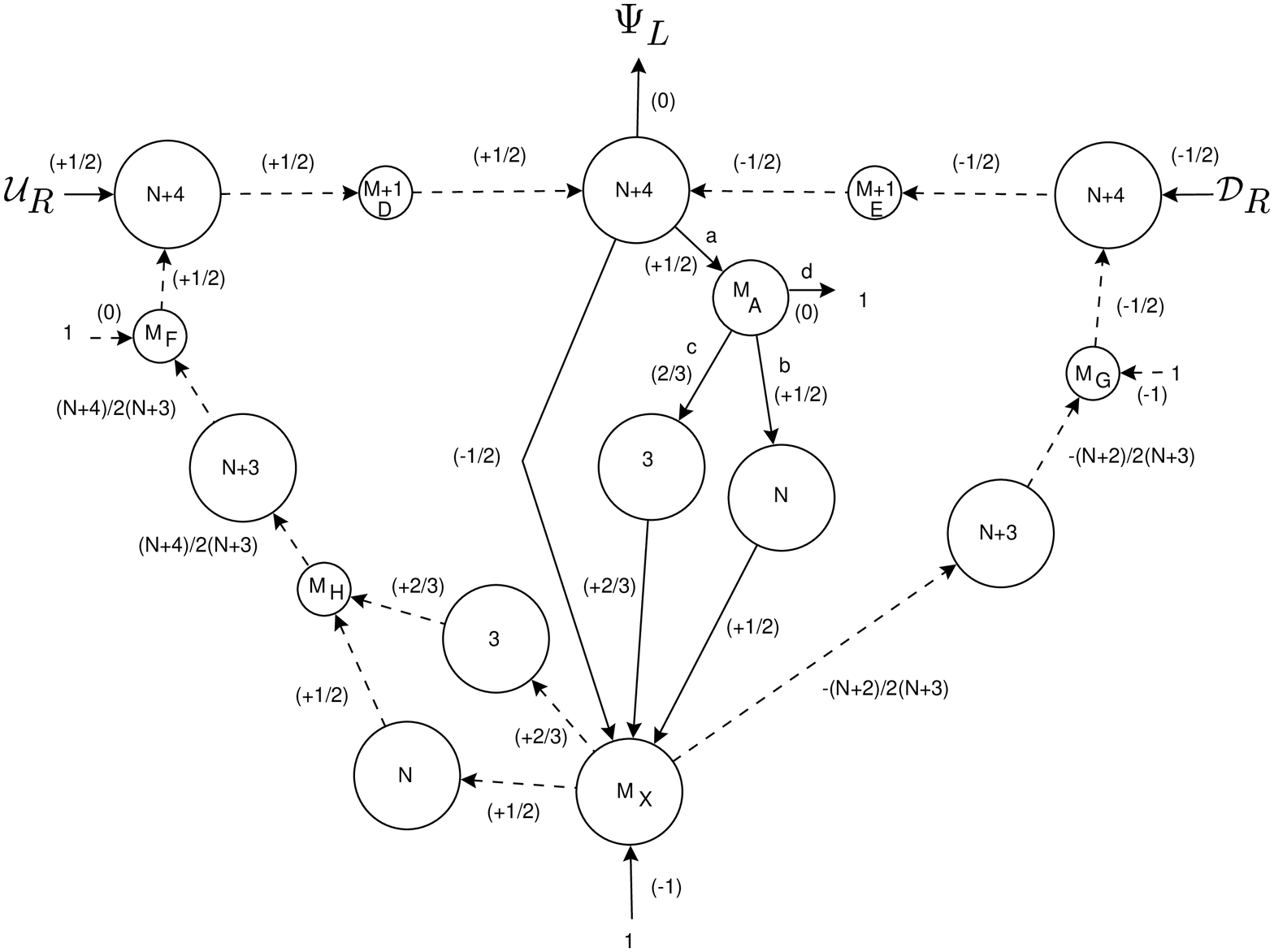}} \vspace{-1.55cm}
\begin{center} Figure 7: Chiral top color in the model of the third family.
\end{center}

We can extend our model to include a top color interaction as shown in Fig 7.
The new $SU(M)_H$ group becomes strongly interacting at $\Lambda_H $ breaking
the  $SU(N+3)_{\cal{U}_R}$ group, left after the right handed neutrino has
decoupled, to $SU(N) \otimes SU(3)$. The right handed SU(3) color group's
coupling will run independently of the technicolor coupling below this breaking
scale (we require that $\Lambda_H$ is large enough that there is enough running
time for the SU(3) and SU(N) groups' couplings to significantly diverge) and
this interaction of the top will be assumed to be near critical when broken to
the vector QCD subgroup  at $\Lambda_D$. Unlike in Hill's model in which the
top bottom mass
splitting is the result of a strongly coupled $U(1)_Y$ gauge interaction (with
the associated problem of it's coupling being close to it's Landau pole) here
the isospin splitting is provided by chiral, asymptotically free, {\it
non-abelian} gauge  groups.

 \section{ Summary and Conclusions}

We have presented a one-doublet technicolor model in which the ETC gauge
symmetries are broken by purely QCD-like dynamics. All fermions transform only
under the fundamental representation of gauge groups. The model has chiral ETC
gauge groups, explicitly breaking  custodial symmetry, and Pati-Salam
unification at high scales. It's main features are:

\begin{itemize}
{\item Three families of quarks and leptons are incorporated, with a hierarchy
of three family-symmetry breaking scales. Within the third family, the full
spectrum of masses can be accommodated. In particular, we argue that with a
third family ETC scale on  the order of $1TeV$,  it may  be possible to
generate both the top and bottom masses through perturbative ETC interactions.
A light tau neutrino mass can be achieved by breaking the ETC group for right
handed isospin $+1/2$ fermions at a high scale. To place $m_{\nu_\tau}$ below
the current limit of roughly 30MeV, this scale must be above about 100TeV.}
{\item Since the model contains a single doublet of technifermions, no
pseudo-Goldstone bosons are formed at the electroweak scale and the $S$
parameter can be kept relatively small. The weak custodial isospin symmetry
breaking built into the model leads to a so called ``direct" contribution
\cite{Tom} to $\Delta \rho \equiv \alpha T$, which is first order in the ETC
interaction. Our naive estimate suggests that this contribution may be nearly
$1\%$ and hence possibly above the experimental limit. A more detailed analysis
of this contribution (and that to the $Z b {\bar b}$ vertex) will be given in a
succeeding paper. The ``indirect" contribution, arising from loops of
non-degenerate technifermions, is second order in ETC interactions and is small
relative to the direct contribution when the ETC interactions are
perturbative.}
{\item The model contains global symmetries at the ETC scales, whose
spontaneous breaking leads to massless Goldstone bosons. They can couple to
ordinary matter through SM interactions and are ruled out by stellar energy
loss constraints \cite{redgiant}. They can, however, be given
phenomenologically acceptable masses by further unifications above the ETC
scales, which break the global symmetries. }
{\item Some of the mass splittings within the first two families will be fed
down naturally from the third family. We have argued that the charm strange
mass splitting may be a result of the top bottom mass splitting. The
suppression of all three generations of neutrino masses may be explained by a
single ETC breaking scale. We have not discussed the origin of quark mixing
angles in this work though it will clearly be important to address this point
in the future.}
{\item We have also  demonstrated that a large top quark mass can be generated
dynamically in technicolor  by a near critical top color interaction without
the need for a strong U(1) interaction. This variant of the model is compatible
with the experimental value of $\Delta \rho$.}
\end{itemize}

The model presented here illustrates that ETC symmetries can be broken using
only  QCD-like dynamics and  fermions in  fundamental representations. The
requisite number of quark-lepton and isospin symmetry violating parameters may
be introduced to accomodate the third family spectrum. It remains to be seen
whether this approach leads to an explanation of quark and lepton masses and
CKM mixing angles. \vspace{1in}

\noindent {\bf Acknowledgements}

The authors would like to thank Steve Hsu, Steve Selipsky, Andy Cohen, Sekhar
Chivukula, Ken Lane, Liz Simmons and Lisa Randall for useful comments and
discussion.

\end{document}